# Improving Compressed Counting

**Ping Li**
Department of Statistical Science
Cornell University
Ithaca NY 14853
pingli@cornell.edu

## Abstract

*Compressed Counting (CC)* [22] was recently proposed for estimating the $\alpha$th frequency moments of data streams, where $0 < \alpha \leq 2$. CC can be used for estimating Shannon entropy, which can be approximated by certain functions of the $\alpha$th frequency moments as $\alpha \to 1$. Monitoring Shannon entropy for anomaly detection (e.g., DDoS attacks) in large networks is an important task.

This paper presents a new algorithm for improving CC. The improvement is most substantial when $\alpha \to 1-$. For example, when $\alpha = 0.99$, the new algorithm reduces the estimation variance roughly by **100-fold**. This new algorithm would make CC considerably more practical for estimating Shannon entropy. Furthermore, the new algorithm is statistically optimal when $\alpha = 0.5$.

## 1 Introduction

Real-world data are often dynamic and can be modeled as data streams [2, 13, 29]. For example, reported in *Information Week* (Jan 9, 2006), Wal-Mart refreshes sales data hourly, adding a billion records of data daily. Ming data streams in (e.g.,) 100 TB scale databases has become an important area of research, e.g., [1, 6], as network data can easily reach that scale [37]. Search engines are a typical source of data streams [2].

### 1.1 Data Stream Models

The *Turnstile* model [29] is popular for data streams. The input stream $a_t = (i_t, I_t)$, $i_t \in [1, D]$ arriving sequentially describes the underlying signal $A$,

$$A_t[i_t] = A_{t-1}[i_t] + I_t, \qquad (1)$$

where the increment $I_t$ can be either positive (insertion) or negative (deletion). This model is particularly useful for describing the empirical distribution because $A_t$ could be viewed as dynamic histograms. In Web and network applications, $D = 2^{64}$ may be possible.

In many applications, the *strict-Turnstile* model suffices, which restricts that $A_t[i] \geq 0$ at all times. For example, in an online store, a user normally can not cancel an order unless she/he did place the order.

**Compressed Counting (CC)** assumes a *relaxed strict-Turnstile* model by enforcing $A_t[i] \geq 0$ only at a particular time $t$. For $s \neq t$, $A_s[i]$ can be arbitrary.

### 1.2 Frequency Moments of Data Streams

This study is mainly concerned with computing (approximating, estimating) the $\alpha$th frequency moment

$$F_{(\alpha)} = \sum_{i=1}^{D} A_t[i]^\alpha,$$

where we drop the subscript $t$ in the notation $F_{t,(\alpha)}$.

Clearly, $F_{(\alpha)}$ can be obtained by using a (very long) vector of length $D$ (e.g., $D = 2^{64}$). Because entries of $A_t[i]$ are frequently updated at high rate, computing $F_{(\alpha)}$ using a small storage space and in one pass of the data is crucial. It is often the case that streaming data are not stored, even on disks [2]. The problem of approximating $F_{(\alpha)}$ has been very heavily studied, e.g., [3, 5, 7, 9, 15, 16, 20, 21, 26, 32, 35].

Interestingly, the sum (e.g., $\alpha = 1$) can be computed trivially using a simple counter, assuming the *relaxed strict-Turnstile* model, because $F_{(1)} = \sum_{i=1}^{D} A_t[i] = \sum_{s=0}^{t} I_s$. Thus, the problem of approximating $F_{(\alpha)}$ around $\alpha = 1$ becomes very interesting from a theoretical point of view. The problem of estimating $F_{(\alpha)}$ around $\alpha = 1$ is also practically important. For example, it is known that Shannon entropy can be estimated from certain functions of $F_{(\alpha)}$ by letting $\alpha \to 1$.

### 1.3 Entropies of Data Streams

Shannon entropy is widely used for characterizing the distribution of data streams, especially in Web and networks [8,18,28,38]. The empirical Shannon entropy is defined as

$$H = -\sum_{i=1}^{D} \frac{A_t[i]}{F_{(1)}} \log \frac{A_t[i]}{F_{(1)}}. \quad (2)$$

It is known that Shannon entropy may be approximated by Rényi entropy [30] or Tsallis entropy [34]. Rényi entropy [30], denoted by $H_\alpha$, is defined as

$$H_\alpha = \frac{1}{1-\alpha} \log \frac{\sum_{i=1}^{D} A_t[i]^\alpha}{\left(\sum_{i=1}^{D} A_t[i]\right)^\alpha} \quad (3)$$

Tsallis entropy [34], denoted by $T_\alpha$, is defined as,

$$T_\alpha = \frac{1}{\alpha-1}\left(1 - \frac{F_{(\alpha)}}{F_{(1)}^\alpha}\right). \quad (4)$$

As $\alpha \to 1$, both Rényi entropy and Tsallis entropy converge to Shannon entropy:

$$\lim_{\alpha \to 1} H_\alpha = \lim_{\alpha \to 1} T_\alpha = H.$$

This fact has been explored by everal studies [11,12,38] to approximate Shannon entropy.

Rényi entropy and Tsallis entropy were not originally proposed as tools for approximating Shannon entropy. Many applications (e.g., Ecology, Theoretical CS, Statistical Physics [14, 27, 31]) used Rényi entropy and Tsallis entropy with $\alpha$'s other than $\alpha \approx 1$.

Thus, while we focus on improving estimates of $F_{(\alpha)}$ as $\alpha \to 1$, it is practically useful if our algorithm could also exhibit improvements for $\alpha$ away from 1.

### 1.4 Application: Anomaly Detection in Large-Scale Networks

Measuring entropies of network traffic data has become an important task for detecting anomalies, which include network failures and distributed denial of service (DDoS) attacks [4, 8, 17, 18, 36, 38].

Anomaly events often change the distribution of network traffic. Therefore, the *Turnstile* model, which may be viewed as histograms, becomes useful in real-time network measurements/monitoring.

A recent study [38] applied *symmetric stable random projections* [15, 21] to estimate the $\alpha$th frequency moments $F_{(\alpha)}$ and the Shannon entropy. One major problem they reported is that, to ensure a sufficient accuracy, the required sample size tend to be prohibitive, for example in the order of $O(10^4)$.

### 1.5 Our Contributions

We provide a high-level description of the algorithm, based on *maximally-skewed stable random projections* [22], also called *Compressed Counting (CC)*.

Conceptually, we can view the data stream $A_t$ as a vector $\in \mathbb{R}^D$ and multiple it by a matrix $\mathbf{R} \in \mathbb{R}^{D \times k}$. The resultant projected vector

$$X = \mathbf{R}^T A_t = (x_1, x_2, ..., x_k) \in \mathbb{R}^k$$

has $k$ "samples" and should be much easier to store if $k$ is small (e.g., $k = 100$), compared to (e.g.,) $D = 2^{64}$.

Of course, the matrix-vector multiplication $X = \mathbf{R}^T A_t$ is conducted incrementally. Recall the *Turnstile* model (1) adopts a linear updating rule and the matrix-vector multiplication is also linear. The matrix $\mathbf{R}$ should not be fully materialized. We only need to re-generate entries of $\mathbf{R}$ on the fly and update the corresponding elements in the stored vector $X = (x_1, x_2, ..., x_k)$. This is the standard technique in data streams [15].

The entries of $\mathbf{R}$ are sampled i.i.d. from a *maximally skewed stable distribution*, i.e., $r_{ij} \sim S(\alpha, \beta, 1)$, which has three parameters: $\alpha$ specifies which $F_{(\alpha)}$ we want to compute, $\beta$ is the skewness parameter and is set to be $\beta = 1$, and the scale parameter is 1.

The next task is to estimate $F_{(\alpha)}$ from $X = (x_1, x_2, ..., x_k)$. We will review that, after projections, $x_j \sim S\left(\alpha, \beta = 1, \sum_{i=1}^{D} A_t[i]^\alpha\right)$. In other words, the scale parameter is the $F_{(\alpha)}$ we are after.

The proposed algorithm (estimator) in this study is based on the *optimal power* of the samples, i.e.,

$$\hat{F}_{(\alpha),op} \propto \left(\sum_{j=1}^{k} |x_j|^{\lambda^* \alpha}\right)^{1/\lambda^*},$$

where $\lambda^*$ is carefully pre-chosen and fixed as a constant for each $\alpha$. It is called the *optimal power* estimator because $\lambda^* = \lambda^*(\alpha)$ is chosen to minimize its asymptotic (as $k \to \infty$) variance at each $\alpha$.

In [22], two estimators were provided, based on the *geometric mean*

$$\hat{F}_{(\alpha),gm} \propto \prod_{j=1}^{k} |x_j|^{1/k}$$

and the *harmonic mean*

$$\hat{F}_{(\alpha),hm} \propto \frac{1}{\sum_{j=1}^{k} |x_j|^{-\alpha}}.$$

When $\alpha \to 1$, these two estimators are far from optimal. In comparison, the *geometric mean* estimator for *symmetric stable random projections* [21] is close to be statistically optimal around $\alpha = 1$.

We can compare the estimators using their asymptotical variances. The left panel of Figure 1 demonstrates that, CC dramatically improves *symmetric stable random projections* as $\alpha \to 1$, and the *optimal power* estimator improves both the *geometric mean* and *harmonic mean* estimators of CC.

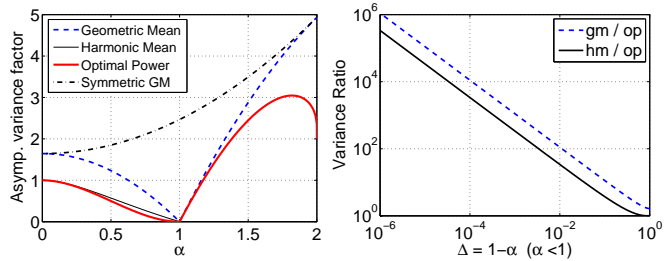

Figure 1: Suppose $\hat{F}$ is an estimator of $F$ with asymptotic variance $\text{Var}\left(\hat{F}\right) = V\frac{F^2}{k} + O\left(\frac{1}{k^2}\right)$. The left panel plots the $V$ values for the *geometric mean* estimator, the *harmonic mean* estimator (for $\alpha < 1$), and the *optimal power* estimator, along with the $V$ values for the *geometric mean* estimator for *symmetric stable random projections* in [21] ("symmetric GM"). The right panel plots the ratios of the variances to better illustrate the significant improvement of the *optimal power* estimator, near $\alpha = 1$ (and $\alpha < 1$).

The improvement of the new estimator $\hat{F}_{(\alpha),op}$ is highly significant when $\alpha \to 1-$. To better illustrate this result, the right panel of Figure 1 plots the ratios of the variances for directly comparing the *geometric mean* and *harmonic mean* estimators with the *optimal power* estimator. For example, when $\Delta = 1 - \alpha \approx 0.01$ (or 0.0001), the variance ratios are about **100** (or 10000). In other words, when $\alpha \approx 0.99$ (or 0.9999), the variance of the *optimal power* estimator is only about 1% (or 0.01%) of the variance of the *geometric mean* (or *harmonic mean*) estimator. This is a remarkable improvement and will be practically highly useful.

### 1.6 The Significance of the Improvement

We would like to introduce the theoretical results [11, 12], which proposed the rigorous criterion on how to choose $\alpha$ so that the Tsallis entropy or Rényi entropy is close enough to Shannon entropy. Their results are based on the worst-case bias. In general, [11, 12] proposed using $\alpha$ extremely close to 1, e.g., $\Delta = |1 - \alpha| < 0.0001$ or even smaller.

Therefore, according to the criteria in [11, 12], our proposed estimator would be very useful; whereas the previous estimators in [22] are not yet practical.

The rest of the paper is organized as follows. The methodologies of CC and two estimators are reviewed in Section 2. The new estimator based on the *optimal power* is presented in Section 3. We show in Section 4 that the *optimal power* estimator is statistically optimal when $\alpha = 0.5$. Some experiments are provided in Section 5 to help demonstrate the remarkable improvements of the new algorithm for CC.

## 2 Review of Compressed Counting

Compressed Counting (CC) is based on *maximally-skewed stable random projections*.

### 2.1 Maximally-Skewed Stable Distributions

A random variable $Z$ is maximally-skewed $\alpha$-stable if the Fourier transform of its density is [39]

$$\mathscr{F}_Z(t) = \text{E}\exp\left(\sqrt{-1}Zt\right)$$
$$= \exp\left(-F|t|^\alpha\left(1 - \sqrt{-1}\beta\text{sign}(t)\tan\left(\frac{\pi\alpha}{2}\right)\right)\right),$$

where $0 < \alpha \leq 2$, $F > 0$, and $\beta = 1$. We denote $Z \sim S(\alpha, \beta = 1, F)$. The skewness parameter $\beta$ for general stable distributions ranges in $[-1, 1]$; but CC uses $\beta = 1$, i.e., **maximally-skewed**. Previously, *symmetric stable random projections* [15, 21] used $\beta = 0$.

Consider independent variables, $Z_1$, $Z_2 \sim S(\alpha, \beta = 1, 1)$. For any constants $C_1$, $C_2 \geq 0$, the "$\alpha$-stability" follows from properties of Fourier transforms:

$$Z = C_1 Z_1 + C_2 Z_2 \sim S\left(\alpha, \beta = 1, C_1^\alpha + C_2^\alpha\right).$$

When $\beta = 0$, the above stability holds even if $C_1$ and $C_2$ are negative. This is why *symmetric stable random projections* [15, 21] can be applied to general data but CC is restricted to the *relaxed strict-Turnstile model*.

### 2.2 Random Projections

Conceptually, one can generate a matrix $\mathbf{R} \in \mathbb{R}^{D \times k}$ and multiply it with the data stream $A_t$, i.e., $X = \mathbf{R}^\text{T} A_t \in \mathbb{R}^k$. The resultant vector $X$ is only of length $k$. The entries of $\mathbf{R}$, $r_{ij}$, are i.i.d. samples of a stable distribution $S(\alpha, \beta = 1, 1)$.

By property of Fourier transforms, the entries of $X$, $x_j$ $j = 1$ to $k$, are i.i.d. samples of a stable distribution

$$x_j = \left[\mathbf{R}^\text{T} A_t\right]_j = \sum_{i=1}^D r_{ij} A_t[i] \sim S\left(\alpha, \beta = 1, \sum_{i=1}^D A_t[i]^\alpha\right),$$

whose scale parameter is exactly $F_{(\alpha)}$. Thus, CC boils down to a statistical estimation problem.

For real implementation, one should conduct $\mathbf{R}^\text{T} A_t$ incrementally. This is possible because the *Turnstile* model (1) is linear. For every incoming $a_t = (i_t, I_t)$, we update $x_j \leftarrow x_j + r_{i_t j} I_t$ for $j = 1$ to $k$. Entries of $\mathbf{R}$ are generated on-demand as necessary.

### 2.3 The Efficiency in Processing Time

Prior to CC, the method of *symmetric stable random projections* was the standard algorithm in data stream computations. However, [9] commented that, when $k$ is large, generating entries of **R** on-demand and multiplications $r_{i_t j} \times I_t$, $j = 1$ to $k$, can be too prohibitive. An easy "fix" is to use $k$ as small as possible, which is possible with CC when $\alpha \approx 1$.

At the same sample size $k$, all procedures of CC and *symmetric stable random projections* are the same except that the entries in **R** follow different distributions. Since CC is much more accurate especially when $\alpha \approx 1$, it requires a much smaller sample size $k$.

### 2.4 Two Statistical Estimators for CC

CC boils down to estimating $F_{(\alpha)}$ from $k$ i.i.d. samples $x_j \sim S(\alpha, \beta = 1, F_{(\alpha)})$. This is an interesting problem because stable distributions in general do not have closed-form density functions. Closed-form density functions exist when $\alpha = 2$ (normal), or $\alpha = 1$ with $\beta = 0$ (Cauchy), or $\alpha = 0.5$ with $\beta = 1$ (Lévy).

[22] provided two statistical estimators.

#### 2.4.1 The Geometric Mean Estimator

$$\hat{F}_{(\alpha),gm} = \frac{\prod_{j=1}^k |x_j|^{\alpha/k}}{D_{gm}}$$

$$D_{gm} = \left( \cos^k\left(\frac{\kappa(\alpha)\pi}{2k}\right) / \cos\left(\frac{\kappa(\alpha)\pi}{2}\right) \right)$$

$$\times \left[ \frac{2}{\pi} \sin\left(\frac{\pi\alpha}{2k}\right) \Gamma\left(1 - \frac{1}{k}\right) \Gamma\left(\frac{\alpha}{k}\right) \right]^k,$$

$$\kappa(\alpha) = \alpha, \quad \text{if } \alpha < 1, \quad \kappa(\alpha) = 2 - \alpha \text{ if } \alpha > 1,$$

which is unbiased and has asymptotic (i.e., as $k \to \infty$) variance

$$\text{Var}\left(\hat{F}_{(\alpha),gm}\right) = \begin{cases} \frac{F_{(\alpha)}^2}{k} \frac{\pi^2}{6}(1 - \alpha^2) + O\left(\frac{1}{k^2}\right), & \alpha < 1 \\ \frac{F_{(\alpha)}^2}{k} \frac{\pi^2}{6}(\alpha - 1)(5 - \alpha) + O\left(\frac{1}{k^2}\right), & \alpha > 1 \end{cases}$$

As $\alpha \to 1$, the asymptotic variance approaches zero.

#### 2.4.2 The Harmonic Mean Estimator ($\alpha < 1$)

$$\hat{F}_{(\alpha),hm} = \frac{k \frac{\cos\left(\frac{\alpha\pi}{2}\right)}{\Gamma(1+\alpha)}}{\sum_{j=1}^k |x_j|^{-\alpha}} \left(1 - \frac{1}{k}\left(\frac{2\Gamma^2(1+\alpha)}{\Gamma(1+2\alpha)} - 1\right)\right),$$

which is asymptotically unbiased and has variance

$$\text{Var}\left(\hat{F}_{(\alpha),hm}\right) = \frac{F_{(\alpha)}^2}{k}\left(\frac{2\Gamma^2(1+\alpha)}{\Gamma(1+2\alpha)} - 1\right) + O\left(\frac{1}{k^2}\right).$$

### 2.5 Previous Estimators Are Inadequate

While the above two estimators are nice results, they are not adequate for estimating Shannon entropy $H$. From the definitions of Rényi entropy $H_\alpha$ (3) and Tsallis entropy $T_\alpha$ (4), we can see immediately that using the estimated $H_\alpha$ and $T_\alpha$ to approximate $H$, the estimation variances would be proportional to $\frac{1}{(\alpha-1)^2}$, which blows up very quickly as $\alpha \to 1$. However, the estimation variance of the *geometric mean* estimator $\hat{F}_{(\alpha),gm}$ decreases to zero at the rate of $O(|\alpha - 1|)$, which is too slow to cancel $\frac{1}{(\alpha-1)^2}$.

According [11, 12], if a high accuracy is required, we have to use $\alpha$ close 1, because the only way to reduce the "intrinsic bias," $|H_\alpha - H|$ or $|T_\alpha - H|$, is to let $\alpha$ be close to 1 as possible. Therefore, we need better algorithms (estimators); and the *optimal power* estimator provides such an algorithm.

## 3 The Optimal Power Estimator

Recall the task is to estimate the scale parameter $F_{(\alpha)}$ from $k$ i.i.d. samples $x_j \sim S(\alpha, \beta = 1, F_{(\alpha)})$. The *optimal power* estimator is based on the fundamental results on the moments of stable distributions.

**Lemma 1** *[22].* *If $Z \sim S(\alpha, \beta = 1, F_{(\alpha)})$, then for any $-1 < \lambda < \alpha$,*

$$\mathbf{E}\left(|Z|^\lambda\right) = F_{(\alpha)}^{\lambda/\alpha} \frac{G(\lambda)}{\cos^{\lambda/\alpha}\left(\frac{\kappa(\alpha)\pi}{2}\right)}, \quad \text{where}$$

$$\kappa(\alpha) = \alpha \quad \text{if} \quad \alpha < 1, \quad \text{and} \quad \kappa(\alpha) = 2 - \alpha \quad \text{if} \quad \alpha > 1.$$

$$G(\lambda) = \frac{2}{\pi} \cos\left(\frac{\kappa(\alpha)}{\alpha}\frac{\lambda\pi}{2}\right) \sin\left(\frac{\pi}{2}\lambda\right) \Gamma\left(1 - \frac{\lambda}{\alpha}\right) \Gamma(\lambda),$$

(5)

*In particular, if $\alpha < 1$, then, for any $-\infty < \lambda < \alpha$,*

$$\mathbf{E}\left(|Z|^\lambda\right) = F_{(\alpha)}^{\lambda/\alpha} \frac{\Gamma\left(1 - \frac{\lambda}{\alpha}\right)}{\cos^{\lambda/\alpha}\left(\frac{\alpha\pi}{2}\right) \Gamma(1 - \lambda)}.$$

The idea of the *optimal power* estimator is to first find an unbiased estimator of $F_{(\alpha)}^\lambda$, which will be in the form of $\frac{1}{k} \sum_{i=1}^k |x_j|^{\lambda^*\alpha}$, using the moment formulas in Lemma 1. Next, we apply $()^{1/\lambda}$ operation to the estimator of $F_{(\alpha)}^\lambda$ to obtain a (biased) estimator of $F_{(\alpha)}$. The $O\left(\frac{1}{k}\right)$ bias can be removed by standard Taylor expansion method (the "Delta" method in statistics).

Lemma 2 presents the *optimal power* estimator, taking into account of bias-correction.

**Lemma 2** *The* optimal power *estimator is*

$$\hat{F}_{(\alpha),op} = \left(\frac{1}{k} \frac{\cos^{\lambda^*}\left(\frac{\kappa(\alpha)\pi}{2}\right) \sum_{j=1}^k |x_j|^{\lambda^*\alpha}}{G(\alpha\lambda^*)}\right)^{1/\lambda^*},$$

where the function $G(\lambda)$ is defined in (5).

With bias-correction, the estimator becomes

$$\hat{F}_{(\alpha),op,c} = \hat{F}_{(\alpha),op}\left\{1 - \frac{1}{k}\frac{1}{2\lambda^*}\left(\frac{1}{\lambda^*}-1\right)\left[\frac{G(2\alpha\lambda^*)}{G^2(\alpha\lambda^*)}-1\right]\right\}$$

whose bias and variance are

$$E\left(\hat{F}_{(\alpha),op,c}\right) = F_{(\alpha)} + O\left(\frac{1}{k^2}\right)$$

$$Var\left(\hat{F}_{(\alpha),op,c}\right) = F_{(\alpha)}^2 \frac{1}{\lambda^{*2}k}\left[\frac{G(2\alpha\lambda^*)}{G^2(\alpha\lambda^*)}-1\right] + O\left(\frac{1}{k^2}\right).$$

The parameter $\lambda^* = \lambda^*(\alpha)$ is determined by solving an optimization problem:

$$\lambda^* = argmin\ g(\lambda;\alpha), \quad where \quad g(\lambda;\alpha) = \frac{1}{\lambda^2}\left[\frac{G(2\alpha\lambda)}{G^2(\alpha\lambda)}-1\right]$$

**Proof:**　　See Appendix A.□

Although the expressions appear very sophisticated, the real computations only involve $\left(\frac{1}{k}\sum_{i=1}^{k}|x_j|^{\lambda^*\alpha}\right)^{1/\lambda^*}$ because all other terms are functions of $\lambda^*$ and $\alpha$ and can be pre-computed.

Figure 2(a) plots $g(\lambda;\alpha)$ in Lemma 2 as functions of $\lambda$ for a good range of $\alpha$ values, illustrating that $g(\lambda;\alpha)$ is a convex function of $\lambda$ and hence the minimums $\lambda^*$ can be easily obtained for every $\alpha$. This fact will be proved for $\alpha < 1$ in Lemma 3.

Examples of the optimal power values $\lambda^* = \lambda^*(\alpha)$: $\lambda^*(0) = -1$, $\lambda^*(0.5) = -2$, $\lambda^*(0.99) = -114.9$.

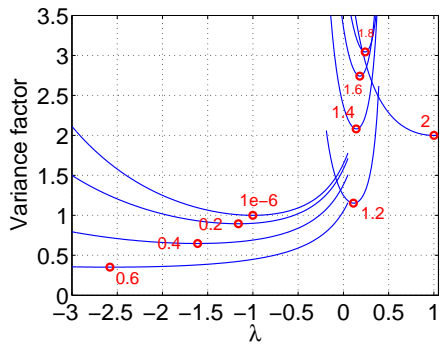

Figure 2: $g(\lambda;\alpha)$ in Lemma 2 as functions of $\lambda$. The number labeled on the lowest point of each curve is $\alpha$.

This type of estimator was also proposed in [25], for *symmetric stable random projections*, which improved the estimators in [21] by merely $0 \sim 30\%$ (depending on $\alpha$). The *optimal power* estimator in [25] had only finite moments to a rather limited order (which seriously affects statistical properties). In comparison, the improvements of the new *optimal power* estimator for CC would be, for example, **100-fold** or more at $\alpha \approx 1$.

The theoretical analysis of this *optimal power* estimator is far more sophisticated, compared to [25]. When $\alpha > 1$, we can see from Figure 1 that the improvement is not very significant (unless $\alpha$ is close to 2). Thus, our theoretical analysis of the *optimal power estimator* mainly focuses on $\alpha < 1$.

For $\alpha < 1$, Lemma 3 proves that the optimal power $\lambda^* < 0$, implying that all moments exist, according to results in Lemma 1. Lemma 3 also proves that $g(\lambda;\alpha)$ is a convex function of $\lambda$ for every $\alpha$ if $\alpha < 1$.

**Lemma 3** *If $\alpha < 1$, then $g(\lambda;\alpha)$ is a convex function of $\lambda$ and the optimal solution $\lambda^* < 0$.*

**Proof:**　　See Appendix B.□

Therefore, if $\alpha < 1$, the *optimal power estimator* has all the moments, suggesting that this estimator may have good statistical properties. Figure 1 (right panel) shows that as $\alpha \to 1-$, the improvements of the *optimal power* over previous estimators in [22] are substantial and will be practically highly useful.

## 4　Optimal Estimator at $\alpha = 0.5$

For the *optimal power* estimator $\hat{F}_{(\alpha),op,c}$, we can verify that when $\alpha = 0.5$, the solution $\lambda = -2$ satisfies $\frac{\partial g(\lambda;\alpha)}{\partial \lambda} = 0$. Because $g(\lambda;\alpha)$ is a convex function, we know $\lambda^* = -2$ when $\alpha = 0.5$. Thus, after simplifying the expression in Lemma 2 for $\alpha = 0.5$, we obtain

$$\hat{F}_{(0.5),op,c} = \left(1 - \frac{3}{4}\frac{1}{k}\right)\sqrt{\frac{k}{\sum_{j=1}^{k}\frac{1}{x_j}}}.$$

It turns out $\hat{F}_{(0.5),op,c}$ is exactly the same as the maximum likelihood estimator (MLE) at $\alpha = 0.5$, with bias-correction, which is statistically optimal in that the asymptotic variance attains the Cramér-Rao lower bound, according to standard statistics results.

When $\alpha = 0.5$ and $\beta = 1$, the stable distribution is known as the *Lévy* distribution, whose the density function can be expressed in a closed-form:

$$f_Z(z) = \frac{F_{(0.5)}}{\sqrt{2\pi}}\frac{\exp\left(-\frac{F_{(0.5)}^2}{2z}\right)}{z^{3/2}}.$$

Lemma 4 derives the MLE and the moments. The proof (which is omitted) involves tedious algebraic work by applying general results in [33].

**Lemma 4** *Assume $x_j \sim S(0.5, 1, F_{(0.5)})$, $j = 1$ to $k$, i.i.d. The maximum likelihood estimator of $F_{(0.5)}$, is*

$$\hat{F}_{(0.5),mle} = \sqrt{\frac{k}{\sum_{j=1}^{k}\frac{1}{x_j}}}.$$

The bias-corrected version is

$$\hat{F}_{(0.5),mle,c} = \left(1 - \frac{3}{4}\frac{1}{k}\right)\sqrt{\frac{k}{\sum_{j=1}^{k} \frac{1}{x_j}}}.$$

The first four moments of $\hat{F}_{(0.5),mle,c}$ are

$$E\left(\hat{F}_{(0.5),mle,c}\right) = F_{(0.5)} + O\left(\frac{1}{k^2}\right),$$

$$Var\left(\hat{F}_{(0.5),mle,c}\right) = \frac{1}{2}\frac{F_{(0.5)}^2}{k} + \frac{9}{8}\frac{F_{(0.5)}^2}{k^2} + O\left(\frac{1}{k^3}\right),$$

$$E\left(\hat{F}_{(0.5),mle,c} - E\left(\hat{F}_{(0.5),mle,c}\right)\right)^3 = \frac{5}{4}\frac{F_{(0.5)}^3}{k^2} + O\left(\frac{1}{k^3}\right),$$

$$E\left(\hat{F}_{(0.5),mle,c} - E\left(\hat{F}_{(0.5),mle,c}\right)\right)^4$$
$$= \frac{3}{4}\frac{F_{(0.5)}^4}{k^2} + \frac{75}{8}\frac{F_{(0.5)}^4}{k^3} + O\left(\frac{1}{k^4}\right).$$

## 5  An Empirical Study

We used the Web crawl data from a chunk of $D = 2^{16}$ pages. That is, the data vector (of length $D$) represents the numbers of occurrences of a word in $D = 2^{16}$ documents. Here, we only use static data instead of real data streams. This is a valid experiment because we just need to compare the estimation accuracy.

We report the experiments for $\Delta = 1 - \alpha$ from 0.1 to $10^{-5}$, and $k = 5, 10, 100, 1000, 4000$. Our experiments show that even with small $k$, the *optimal power* estimator achieved good accuracies, in terms of the normalized mean square errors ($\sqrt{\text{MSE}}$, divided by the true value). We notice that the normalized MSEs are quite similar across different words. Thus, we only present the results for one word vector, for the word "RICE."

Figure 3 presents the results on estimating frequency moments using both CC and *symmetric stable random projections*. As expected, CC considerably improves *symmetric stable random projections* and the *optimal power* estimator of CC considerably improves the *geometric mean* and *harmonic mean* estimators of CC. Also, the empirical MSEs matched the theoretical asymptotic variances well.

Figure 4 presents the results on estimating Shannon entropy using the estimated Tsallis entropy. Again, we can see the huge improvements of CC over *symmetric stable random projections* and the improvements of *optimal power* estimator over the other two estimators of CC. When $\alpha$ is too close 1, the *geometric mean* and *harmonic mean* estimators exhibit very large errors because the variances grow up quickly as $\alpha \to 1$.

For example, suppose the required accuracy is normalized MSE< 10%. Using the *optimal power* estimator, it suffices to use only $k = 10$ samples, for this data vector. However, using the *geometric mean* or the *harmonic mean* estimator, we should choose $k = 1000$ and a carefully chosen $\alpha$ because if $\alpha$ is too close to 1, the errors become very large (the bias-variance trade-off).

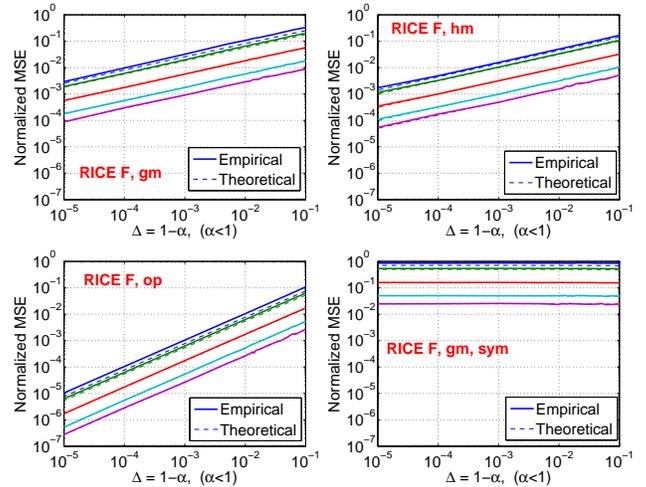

Figure 3: Frequency moments, $F_{(\alpha)}$, for word RICE. The four panels present results of the *geometric mean* (gm) estimator, the *harmonic mean* estimator (hm), the *optimal power* (op) estimator, and the *geometric mean* estimator for *symmetric stable random projections* (gm, sym.). In each panel, each curve represents the normalized MSEs ($\sqrt{\text{MSE}}$, divided by the true value) for a sample size $k$, where $k = 5, 10, 100, 1000$ and 4000. As expected, a curve for smaller $k$ is higher in the panel (i.e., larger MSE).

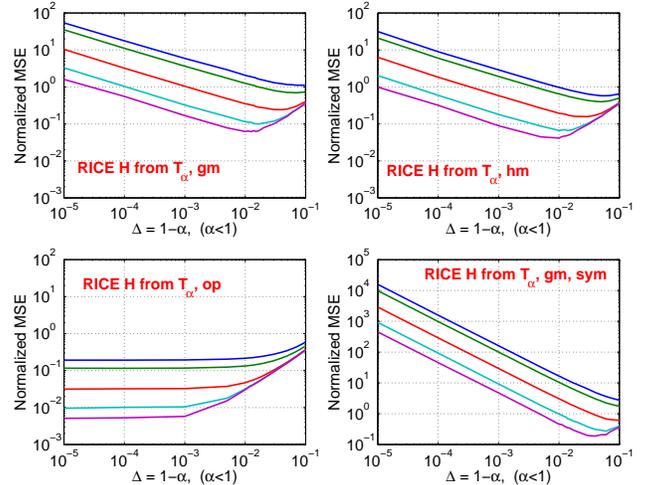

Figure 4: Shannon entropy $H$, estimated from Tsallis entropy $T_\alpha$, for word RICE. In each panel, each curve represents the normalized MSEs ($\sqrt{\text{MSE}}$, divided by the true value) for a sample size $k$, where $k = 5, 10, 100, 1000$ and 4000. A curve for smaller $k$ is higher in the panel (i.e., larger MSE).

Suppose we instead use *symmetric stable random pro-*

*jections*. We need to use a much larger $k$ and very carefully select the $\alpha$. If $\alpha = 0.99$, even with $k = 1000$, the normalized MSE is 1.0, a very large error. Because the $\alpha$ which attains the smallest MSE is data-dependent, this optimal $\alpha$ is generally unknown and hence one may have to use very large $k$ to ensure the accuracy.

Therefore, using CC with the *optimal power* estimator becomes particularly appealing, because one does not have to worry much about choosing the "best" $\alpha$. In this example, $0.99 \leq \alpha < 1$ produced very stable estimates of the Shannon entropy.

There is still room for further substantial improvement. Figure 4 shows that, if a high accuracy is desired (e.g., normalized MSE $< 1\%$), then a large $k$ has to be used (e.g., $k = 1000$) even with the *optimal power* estimator. Figure 4 demonstrates that as $\Delta$ becomes very small (e.g., $\Delta < 10^{-3}$ at $k = 1000$), the errors of the *optimal power* estimator decrease very slowly.

## 6 Conclusions

Data streams are common in real-world. Estimating frequency moments and entropies of data streams is an active area of research. Shannon entropy is important in many applications, for example, real-time mining/monitoring anomaly events (such as DDoS attacks) in large-scale networks. The performance of recent algorithms for estimating Shannon entropy is critically contingent on the quality of the estimated $\alpha$th frequency moments near $\alpha = 1$.

The main contribution is a new algorithm for improving *Compressed Counting (CC)*, which was recently developed for approximating the $\alpha$th frequency moments. Based on the *optimal power* of the samples maintained by CC, our algorithm is as simple as the previous algorithms in [22]. The improvements of the new algorithm is remarkable when $\alpha \to 1-$. For example, in terms of the variances, the improvement would be roughly **100-fold**, when $\alpha = 0.99$; and even larger improvements as $\alpha$ is more close to $1-$.

While the theoretical analysis of the *optimal power* estimator is interesting, this new algorithm will have a practical impact in estimating entropy of data streams.

There are open problems. For example, *what is the true sample complexity bound of CC?* Also, our experiments demonstrate that if a high accuracy is desired, then we still need a large sample size $k$. Therefore, there is much need for future research.

Finally, as a reviewer pointed out, methods based on stable random projections have a limitation. If we need to compute the $\alpha$th moments for 10 different $\alpha$ values, we have to conduct 10 different random projections and store 10 sets of samples. This is the price to pay in order to have algorithms with guaranteed accuracy regardless of the data distribution. In real applications, we sometimes have prior information about the data distribution. For example, large-scale data are often highly sparse. A line of algorithms named *Conditional Random Sampling (CRS)* [23, 24] can be effective in processing sparse data.

## Acknowledgement

Ping Li is partially supported by NSF (DMS-0808864), ONR (N000140910911, YIP), and a gift from Google.

## A  Proof of Lemma 2

We use the $G$ function in (5) to simplify the algebra:

$$G(\lambda) = \frac{2}{\pi} \cos\left(\frac{\kappa(\alpha)}{\alpha} \frac{\lambda \pi}{2}\right) \sin\left(\frac{\pi}{2}\lambda\right) \Gamma\left(1 - \frac{\lambda}{\alpha}\right) \Gamma(\lambda).$$

Assume $x_j \sim S(\alpha, \beta = 1, F_{(\alpha)})$, $j = 1$ to $k$, i.i.d. We first consider $\hat{R}_{(\alpha),\lambda}$, an unbiased estimator of $F_{(\alpha)}^\lambda$:

$$\hat{R}_{(\alpha),\lambda} = \frac{1}{k} \frac{\left|\cos\left(\frac{\alpha\pi}{2}\right)\right|^\lambda \sum_{j=1}^{k} |x_j|^{\lambda\alpha}}{G(\alpha\lambda)},$$

whose variance is

$$\mathrm{Var}\left(\hat{R}_{(\alpha),\lambda}\right) = \frac{F_{(\alpha)}^{2\lambda}}{k} \left[\frac{G(2\alpha\lambda)}{G^2(\alpha\lambda)} - 1\right].$$

To have bounded variances, we need to restrict $-1/2\alpha < \lambda < 1/2$ if $\alpha > 1$, and $\lambda < 1/2$ if $\alpha < 1$.

A biased estimator of $F_{(\alpha)}$ would be $\left(\hat{R}_{(\alpha),\lambda}\right)^{1/\lambda}$, which has $O\left(\frac{1}{k}\right)$ bias. This bias can be removed to an extent by Taylor expansions [19, Theorem 6.1.1]. We call this new estimator the "fractional power" estimator:

$$\hat{F}_{(\alpha),fp,c,\lambda} = \left(\hat{R}_{(\alpha),\lambda}\right)^{1/\lambda} - \frac{\mathrm{Var}\left(\hat{R}_{(\alpha),\lambda}\right)}{2} \frac{1}{\lambda}\left(\frac{1}{\lambda} - 1\right) \left(d_{(\alpha)}^\lambda\right)^{1/\lambda - 2} =$$

$$\left(\frac{1}{k} \frac{\left|\cos\left(\frac{\alpha\pi}{2}\right)\right|^\lambda \sum_{j=1}^{k} |x_j|^{\lambda\alpha}}{G(\alpha\lambda)}\right)^{1/\lambda} \left(1 - \frac{1}{k}\frac{1}{2\lambda}\left(\frac{1}{\lambda} - 1\right)\left[\frac{G(2\alpha\lambda)}{G^2(\alpha\lambda)} - 1\right]\right)$$

where we plug in the estimated $F_{(\alpha)}^\lambda$. The asymptotic variance, $\mathrm{Var}\left(\hat{F}_{(\alpha),fp,c,\lambda}\right)$, would be

$$\mathrm{Var}\left(\hat{R}_{(\alpha),c,\lambda}\right)\left(\frac{1}{\lambda}\left(F_{(\alpha)}^\lambda\right)^{1/\lambda - 1}\right)^2 + O\left(\frac{1}{k^2}\right)$$

$$= F_{(\alpha)}^2 \frac{1}{\lambda^2 k}\left[\frac{G(2\alpha\lambda)}{G^2(\alpha\lambda)} - 1\right] + O\left(\frac{1}{k^2}\right).$$

The optimal $\lambda$, denoted by $\lambda^*$, is then

$$\lambda^* = \underset{\lambda}{\mathrm{argmin}}\left\{\frac{1}{\lambda^2}\left[\frac{G(2\alpha\lambda)}{G^2(\alpha\lambda)} - 1\right]\right\}.$$

We denote by $\hat{F}_{(\alpha),op,c}$ the optimal fractional power estimator $\hat{F}_{(\alpha),fp,c,\lambda^*}$.

## B  Proof of Lemma 3

We consider only $\alpha < 1$, i.e., $\kappa(\alpha) = \alpha$, To prove that

$$g(\lambda; \alpha) = \frac{1}{\lambda^2} \left( \frac{\cos(\kappa(\alpha)\lambda\pi) \frac{2}{\pi}\Gamma(1-2\lambda)\Gamma(2\lambda\alpha)\sin(\pi\lambda\alpha)}{\left[\cos(\kappa(\alpha)\frac{\lambda\pi}{2})\frac{2}{\pi}\Gamma(1-\lambda)\Gamma(\lambda\alpha)\sin\left(\frac{\pi}{2}\lambda\alpha\right)\right]^2} - 1 \right)$$

is a convex function of $\lambda$, where $\lambda < 1/2$, it suffices to show that $\frac{\partial^2 g(\lambda; \alpha)}{\partial \lambda^2} > 0$. Here unless we specify $\lambda = 0$, we always assume $\lambda \neq 0$ to avoid triviality. (It is easy to show $\frac{\partial^2 g(\lambda; \alpha)}{\partial \lambda^2} \to 0$ when $\lambda \to 0$.)

Using Euler's reflection formula $\Gamma(z)\Gamma(1-z) = \frac{\pi}{\sin \pi z}$, we simplify $g(\lambda; \alpha)$ to be

$$g(\lambda; \alpha)$$
$$= \frac{1}{\lambda^2}\left(\frac{\Gamma(1-2\lambda)\Gamma^2(1-\lambda\alpha)}{\Gamma(1-2\lambda\alpha)\Gamma^2(1-\lambda)} - 1\right) = \frac{1}{\lambda^2}\left(\alpha \frac{\Gamma(-2\lambda)\Gamma^2(-\lambda\alpha)}{\Gamma(-2\lambda\alpha)\Gamma^2(-\lambda)} - 1\right)$$
$$= \frac{1}{\lambda^2}\left(\alpha 2^{2\lambda\alpha-2\lambda}\frac{\Gamma(-\lambda+1/2)\Gamma(-\lambda\alpha)}{\Gamma(-\lambda\alpha+1/2)\Gamma(-\lambda)} - 1\right)$$
$$= \frac{1}{\lambda^2}\left(\alpha 2^{2\lambda\alpha-2\lambda} \prod_{s=0}^{\infty}\left[\left(1 + \frac{1/2}{-\lambda\alpha + s}\right)\left(1 - \frac{1/2}{-\lambda + 1/2 + s}\right)\right] - 1\right)$$
$$= \frac{1}{\lambda^2}\left(\alpha 2^{2\lambda\alpha-2\lambda}\prod_{s=0}^{\infty}\frac{(2s-2\lambda\alpha+1)(s-\lambda)}{(s-\lambda\alpha)(2s+1-2\lambda)} - 1\right) = \frac{1}{\lambda^2}(CM-1),$$

where

$$C = C(\lambda; \alpha) = \alpha 2^{2\lambda\alpha - 2\lambda},$$
$$M = M(\lambda; \alpha) = \prod_{s=0}^{\infty} f_s(\lambda; \alpha), \quad f_s(\lambda; \alpha) = \frac{(2s-2\lambda\alpha+1)(s-\lambda)}{(s-\lambda\alpha)(2s+1-2\lambda)},$$

and we have used properties of the Gamma function [10, 8.335.1, 8.325.1]:

$$\Gamma(2z) = \frac{2^{2z-1}}{\sqrt{\pi}}\Gamma(z)\Gamma(z+1/2),$$
$$\frac{\Gamma(\alpha)\Gamma(\beta)}{\Gamma(\alpha+\gamma)\Gamma(\beta-\gamma)} = \prod_{s=0}^{\infty}\left[\left(1+\frac{\gamma}{\alpha+s}\right)\left(1-\frac{\gamma}{\beta+s}\right)\right].$$

With respect to $\lambda$, the first two derivatives of $g(\lambda; \alpha)$ are (denoting $w = \log(2)(2\alpha - 2)$)

$$\frac{\partial g}{\partial \lambda} = \frac{1}{\lambda^2}\left(-\frac{2}{\lambda}(CM-1) + \left(w + \sum_{s=0}^{\infty}\frac{\partial \log f_s}{\partial \lambda}\right)CM\right), \quad \frac{\partial^2 g}{\partial \lambda^2} = -\frac{6}{\lambda^4} + \frac{CM}{\lambda^2} \times$$
$$\left(\frac{6}{\lambda^2} + \sum_{s=0}^{\infty}\frac{\partial^2 \log f_s}{\partial \lambda^2} + \left(w + \sum_{s=0}^{\infty}\frac{\partial \log f_s}{\partial \lambda}\right)^2 - \frac{4}{\lambda}\left(w + \sum_{s=0}^{\infty}\frac{\partial \log f_s}{\partial \lambda}\right)\right).$$

To show $\frac{\partial^2 g}{\partial \lambda^2} > 0$, it suffices to show

$$\frac{\partial^2 g}{\partial \lambda^2}\lambda^4 = 6(CM-1) + CM\lambda^2 \times$$
$$\left(\sum_{s=0}^{\infty}\frac{\partial^2 \log f_s}{\partial \lambda^2} + \left(w + \sum_{s=0}^{\infty}\frac{\partial \log f_s}{\partial \lambda}\right)^2 - \frac{4}{\lambda}\left(w + \sum_{s=0}^{\infty}\frac{\partial \log f_s}{\partial \lambda}\right)\right) > 0.$$

Since $(CM)|_{\lambda=0} = 1$, $(CM)|_{\lambda \neq 0} > 1$ (which is intuitive and will be shown by algebra), it suffices to show

$$T_1(\lambda; \alpha) = 6(CM-1) + \lambda^2 \sum_{s=0}^{\infty}\frac{\partial^2 \log f_s}{\partial \lambda^2}$$
$$+ \lambda^2 \left(w + \sum_{s=0}^{\infty}\frac{\partial \log f_s}{\partial \lambda}\right)^2 - 4\lambda\left(w + \sum_{s=0}^{\infty}\frac{\partial \log f_s}{\partial \lambda}\right) > 0.$$

Since $T_1(0; \alpha) = 0$, it suffices to show $\lambda \frac{\partial T_1}{\partial \lambda} > 0$, where

$$\frac{\partial T_1}{\partial \lambda}$$
$$= (6CM-4)\left(w + \sum_{s=0}^{\infty}\frac{\partial \log f_s}{\partial \lambda}\right) - 2\lambda \sum_{s=0}^{\infty}\frac{\partial^2 \log f_s}{\partial \lambda^2} + \lambda^2 \sum_{s=0}^{\infty}\frac{\partial^3 \log f_s}{\partial \lambda^3}$$
$$+ 2\lambda\left(w + \sum_{s=0}^{\infty}\frac{\partial \log f_s}{\partial \lambda}\right)^2 + 2\lambda^2\left(w + \sum_{s=0}^{\infty}\frac{\partial \log f_s}{\partial \lambda}\right)\sum_{s=0}^{\infty}\frac{\partial^2 \log f_s}{\partial \lambda^2},$$
$$\lambda \frac{\partial T_1}{\partial \lambda} = (6CM-4)\lambda\left(w + \sum_{s=0}^{\infty}\frac{\partial \log f_s}{\partial \lambda}\right)$$
$$- 2\lambda^2 \sum_{s=0}^{\infty}\frac{\partial^2 \log f_s}{\partial \lambda^2} + \lambda^3 \sum_{s=0}^{\infty}\frac{\partial^3 \log f_s}{\partial \lambda^3}$$
$$+ 2\lambda^2\left(w + \sum_{s=0}^{\infty}\frac{\partial \log f_s}{\partial \lambda}\right)^2 + 2\lambda^3\left(w + \sum_{s=0}^{\infty}\frac{\partial \log f_s}{\partial \lambda}\right)\sum_{s=0}^{\infty}\frac{\partial^2 \log f_s}{\partial \lambda^2}$$

Because $CM > 1$ and we will soon show $\lambda\left(w + \sum_{s=0}^{\infty}\frac{\partial \log f_s}{\partial \lambda}\right) > 0$, it suffices to show

$$2\lambda\left(w + \sum_{s=0}^{\infty}\frac{\partial \log f_s}{\partial \lambda}\right) - 2\lambda^2 \sum_{s=0}^{\infty}\frac{\partial^2 \log f_s}{\partial \lambda^2} + \lambda^3 \sum_{s=0}^{\infty}\frac{\partial^3 \log f_s}{\partial \lambda^3}$$
$$+ 2\lambda^2\left(w + \sum_{s=0}^{\infty}\frac{\partial \log f_s}{\partial \lambda}\right)^2 + 2\lambda^3\left(w + \sum_{s=0}^{\infty}\frac{\partial \log f_s}{\partial \lambda}\right)\sum_{s=0}^{\infty}\frac{\partial^2 \log f_s}{\partial \lambda^2}$$
$$= \lambda T_2(\lambda; \alpha) > 0,$$

for which it suffices to show $T_2(0; \alpha) = 0$, and

$$\frac{\partial T_2}{\partial \lambda} = \lambda^2 \sum_{s=0}^{\infty}\frac{\partial^4 \log f_s}{\partial \lambda^4} + 2\left(w + \sum_{s=0}^{\infty}\frac{\partial \log f_s}{\partial \lambda}\right)^2$$
$$+ 8\lambda\left(w + \sum_{s=0}^{\infty}\frac{\partial \log f_s}{\partial \lambda}\right)\sum_{s=0}^{\infty}\frac{\partial^2 \log f_s}{\partial \lambda^2} + 2\lambda^2\left(\sum_{s=0}^{\infty}\frac{\partial^2 \log f_s}{\partial \lambda^2}\right)^2$$
$$+ 2\lambda^2\left(w + \sum_{s=0}^{\infty}\frac{\partial \log f_s}{\partial \lambda}\right)\sum_{s=0}^{\infty}\frac{\partial^3 \log f_s}{\partial \lambda^3} > 0.$$

To this end, we know in order to prove the convexity of $g(\lambda; \alpha)$, it suffices to prove the following:

$$(CM)|_{\lambda=0} = 1, \quad (CM)|_{\lambda \neq 0} > 1, \quad \lambda\left(w + \sum_{s=0}^{\infty}\frac{\partial \log f_s}{\partial \lambda}\right) > 0,$$
$$\sum_{s=0}^{\infty}\frac{\partial^2 \log f_s}{\partial \lambda^2} > 0, \quad \sum_{s=0}^{\infty}\frac{\partial^4 \log f_s}{\partial \lambda^4} > 0,$$
$$4\sum_{s=0}^{\infty}\frac{\partial^2 \log f_s}{\partial \lambda^2} + \lambda \sum_{s=0}^{\infty}\frac{\partial^3 \log f_s}{\partial \lambda^3} > 0,$$

where

$$\sum_{s=0}^{\infty}\frac{\partial \log f_s}{\partial \lambda} = \sum_{s=0}^{\infty}\left(\frac{-2\alpha}{2s-2\lambda\alpha+1} - \frac{1}{s-\lambda} + \frac{\alpha}{s-\lambda\alpha} + \frac{2}{2s+1-2\lambda}\right)$$
$$\sum_{s=0}^{\infty}\frac{\partial^2 \log f_s}{\partial \lambda^2}$$
$$= \sum_{s=0}^{\infty}\left(\frac{-4\alpha^2}{(2s-2\lambda\alpha+1)^2} - \frac{1}{(s-\lambda)^2} + \frac{\alpha^2}{(s-\lambda\alpha)^2} + \frac{4}{(2s+1-2\lambda)^2}\right),$$
$$\sum_{s=0}^{\infty}\frac{\partial^3 \log f_s}{\partial \lambda^3}$$
$$= \sum_{s=0}^{\infty}\left(\frac{-16\alpha^3}{(2s-2\lambda\alpha+1)^3} - \frac{2}{(s-\lambda)^3} + \frac{2\alpha^3}{(s-\lambda\alpha)^3} + \frac{16}{(2s+1-2\lambda)^3}\right),$$
$$\sum_{s=0}^{\infty}\frac{\partial^4 \log f_s}{\partial \lambda^4}$$
$$= \sum_{s=0}^{\infty}\left(\frac{-96\alpha^4}{(2s-2\lambda\alpha+1)^4} - \frac{6}{(s-\lambda)^4} + \frac{6\alpha^4}{(s-\lambda\alpha)^4} + \frac{96}{(2s+1-2\lambda)^4}\right).$$

First, we can show

$$(CM)|_{\lambda=0} = 1, \quad \text{and} \quad \left(w + \sum_{s=0}^{\infty}\frac{\partial \log f_s}{\partial \lambda}\right)|_{\lambda=0} = 0,$$

because

$$CM|_{\lambda=0} = \alpha \lim_{\lambda \to 0} \frac{(1)(-\lambda)}{(-\lambda\alpha)(1)} \prod_{s=1}^{\infty} \frac{(2s+1)(s)}{(s)(2s+1)} = 1, \text{ and}$$

$$\sum_{s=0}^{\infty} \frac{\partial \log f_s}{\partial \lambda}\bigg|_{\lambda=0} = -2\alpha + 2 + \sum_{s=1}^{\infty} \left( \frac{-2\alpha}{2s+1} - \frac{1}{s} + \frac{\alpha}{s} + \frac{2}{2s+1} \right)$$

$$= -2\alpha + 2 + (\alpha - 1) \sum_{s=1}^{\infty} \frac{1}{s(2s+1)} = -(\alpha - 1)\log(2) = -w,$$

using [10, 0.234.8] $\sum_{s=1}^{\infty} \frac{1}{s(2s+1)} = 2 - 2\log(2)$. Therefore, once we have proved $\sum_{s=0}^{\infty} \frac{\partial^2 \log f_s}{\partial \lambda^2} > 0$, it follows that $(CM)|_{\lambda \neq 0} > 1$ and $\lambda \left( w + \sum_{s=0}^{\infty} \frac{\partial \log f_s}{\partial \lambda} \right) > 0$.

To show

$$\sum_{s=0}^{\infty} \frac{\partial^2 \log f_s}{\partial \lambda^2} > 0, \qquad \sum_{s=0}^{\infty} \frac{\partial^4 \log f_s}{\partial \lambda^4} > 0,$$

$$4 \sum_{s=0}^{\infty} \frac{\partial^2 \log f_s}{\partial \lambda^2} + \lambda \sum_{s=0}^{\infty} \frac{\partial^3 \log f_s}{\partial \lambda^3} > 0,$$

we use Riemanns' Zeta function [10, 9.511,9.521],

$$\zeta(m, q) = \sum_{s=0}^{\infty} \frac{1}{(s+q)^m} = \frac{1}{\Gamma(m)} \int_0^{\infty} \frac{t^{m-1} e^{-qt}}{1 - e^{-t}} dt, \quad q < 0, \quad m > 1,$$

to rewrite

$$\sum_{s=0}^{\infty} \frac{\partial^2 \log f_s}{\partial \lambda^2}$$

$$= \sum_{s=0}^{\infty} \left( \frac{-4\alpha^2}{(2s - 2\lambda\alpha + 1)^2} - \frac{1}{(s - \lambda)^2} + \frac{\alpha^2}{(s - \lambda\alpha)^2} + \frac{4}{(2s + 1 - 2\lambda)^2} \right)$$

$$= -\alpha^2 \zeta\left(2, \frac{1}{2} - \lambda\alpha\right) - \frac{1}{\lambda^2} - \zeta(2, 1 - \lambda) + \frac{\alpha^2}{\lambda^2 \alpha^2} + \alpha^2 \zeta(2, 1 - \lambda\alpha) + \zeta\left(2, \frac{1}{2} - \lambda\right)$$

$$= \int_0^{\infty} \frac{t}{1 - e^{-t}} \left( -\alpha^2 \exp(-t(1/2 - \lambda\alpha)) - \exp(-t(1 - \lambda)) \right)$$

$$\quad + \frac{t}{1 - e^{-t}} \left( \alpha^2 \exp(-t(1 - \lambda\alpha)) + \exp(-t(1/2 - \lambda)) \right) dt$$

$$= \int_0^{\infty} \frac{t}{1 - e^{-t}} \left( e^{-t/2} - e^{-t} \right) \left( e^{\lambda t} - \alpha^2 e^{\lambda \alpha t} \right) dt$$

$$= \int_0^{\infty} \frac{t e^{-t/2}}{1 + e^{-t/2}} \left( e^{\lambda t} - \alpha^2 e^{\lambda \alpha t} \right) dt$$

$$= \int_0^{\infty} \frac{t}{1 + e^{-t/2}} \left( e^{-t(1/2 - \lambda)} - \alpha^2 e^{-t(1/2 - \lambda\alpha)} \right) dt$$

Note that $1 \leq 1 + e^{-t/2} \leq 2$ when $t \in [0, \infty)$, and

$$\int_0^{\infty} t \left( e^{-t(1/2 - \lambda)} - \alpha^2 e^{-t(1/2 - \lambda\alpha)} \right) dt$$

$$= \frac{1}{(1/2 - \lambda)^2} - \frac{\alpha^2}{(1/2 - \lambda\alpha)^2} = \frac{1}{(1/2 - \lambda)^2} - \frac{1}{(1/2/\alpha - \lambda)^2} > 0$$

because $\lambda < 1/2$, $\alpha < 1$, and $\int_0^{\infty} t^m e^{-pt} dt = m! p^{-m-1}$. This proves that $\sum_{s=0}^{\infty} \frac{\partial^2 \log f_s}{\partial \lambda^2} > 0$.

Similarly,

$$\sum_{s=0}^{\infty} \frac{\partial^4 \log f_s}{\partial \lambda^4}$$

$$= \sum_{s=0}^{\infty} \left( \frac{-96\alpha^4}{(2s - 2\lambda\alpha + 1)^4} - \frac{6}{(s - \lambda)^4} + \frac{6\alpha^4}{(s - \lambda\alpha)^4} + \frac{96}{(2s + 1 - 2\lambda)^4} \right)$$

$$= -6\alpha^4 \zeta\left(4, \frac{1}{2} - \lambda\alpha\right) - \frac{6}{\lambda^4} - \zeta(4, 1 - \lambda) + \frac{6\alpha^4}{\lambda^4 \alpha^4} + 6\alpha^2 \zeta(4, 1 - \lambda\alpha)$$

$$\quad + 6\zeta\left(4, \frac{1}{2} - \lambda\right) = \int_0^{\infty} \frac{t^3}{1 + e^{-t/2}} \left( e^{-t(1/2 - \lambda)} - \alpha^4 e^{-t(1/2 - \lambda\alpha)} \right) dt$$

$$\geq \frac{3!}{2} \left( \frac{1}{(1/2 - \lambda)^4} - \frac{\alpha^4}{(1/2 - \lambda\alpha)^4} \right) > 0.$$

At this point, it is trivial to show

$$4 \sum_{s=0}^{\infty} \frac{\partial^2 \log f_s}{\partial \lambda^2} + \lambda \sum_{s=0}^{\infty} \frac{\partial^3 \log f_s}{\partial \lambda^3} > 0 \quad \text{if } \lambda > 0.$$

For $\lambda < 0$, however, we need a different approach. Note that when $\alpha \to 1$,

$$W = 4 \sum_{s=0}^{\infty} \frac{\partial^2 \log f_s}{\partial \lambda^2} + \lambda \sum_{s=0}^{\infty} \frac{\partial^3 \log f_s}{\partial \lambda^3} \to 0.$$

Therefore, we can treat $W$ as a function of $\lambda$ for fixed $\lambda$. The only thing we need to show is $\frac{\partial W}{\partial \alpha} < 0$ when $\alpha < 1$ and $\lambda < 0$.

$$\frac{\partial W}{\partial \alpha} = \frac{\partial \left[ 4 \sum_{s=0}^{\infty} \frac{\partial^2 \log f_s}{\partial \lambda^2} + \lambda \sum_{s=0}^{\infty} \frac{\partial^3 \log f_s}{\partial \lambda^3} \right]}{\partial \alpha}$$

$$= \int_0^{\infty} \frac{e^{-t(1/2 - \lambda\alpha)}}{1 + e^{-t/2}} \left( 4t \left[ -2\alpha - \alpha^2 \lambda t \right] + \lambda t^2 \left[ -3\alpha^2 - \alpha^3 \lambda t \right] \right) dt$$

$$= -\int_0^{\infty} \frac{e^{-t(1/2 - \lambda\alpha)}}{1 + e^{-t/2}} \left( 8\alpha t + 7\alpha^2 \lambda t^2 + \alpha^3 \lambda^2 t^3 \right) dt$$

$$\leq -\frac{1}{2} \int_0^{\infty} e^{-t(1/2 - \lambda\alpha)} \left( 8\alpha t + 7\alpha^2 \lambda t^2 + \alpha^3 \lambda^2 t^3 \right) dt$$

$$= -\left( \frac{4\alpha}{(1/2 - \lambda\alpha)} + \frac{7/2\alpha^2 \lambda}{(1/2 - \lambda\alpha)^2} + \frac{\alpha^3 \lambda^2}{(1/2 - \lambda\alpha)^3} \right)$$

$$= -\frac{\alpha}{(1/2 - \lambda\alpha)^3} \left( 4(1/2 - \lambda\alpha)^2 + 7/2\alpha\lambda(1/2 - \lambda\alpha) + \alpha^2 \lambda^2 \right)$$

$$= \frac{-\alpha}{(1/2 - \lambda\alpha)^3} \left( \left[ \alpha\lambda + \frac{7}{4}\left(\frac{1}{2} - \alpha\lambda\right) \right]^2 + \left( 4 - \frac{7^2}{4^2} \right) \left( \frac{1}{2} - \alpha\lambda \right)^2 \right)$$

$$< 0.$$

This completes the proof of the convexity of $g(\lambda; \alpha)$. Finally, we need to show that $\lambda^* < 0$, where $\lambda^*$ is the solution to $\frac{\partial g}{\partial \lambda} = 0$, or equivalently, the solution to

$$V(\lambda; \alpha) = -2(CM - 1) + \lambda \left( w + \sum_{s=0}^{\infty} \frac{\partial \log f_s}{\partial \lambda} \right) CM = 0,$$

provided we discard the trivial solution $\lambda = 0$. Thus, it suffices to show that $V(\lambda; \alpha)$ increases monotonically as $\lambda > 0$, i.e., $\frac{\partial V}{\partial \lambda} > 0$ if $\lambda > 0$. Because

$$\frac{\partial V}{\partial \lambda}$$

$$= CM \left( 2 \left( w + \sum_{s=0}^{\infty} \frac{\partial \log f_s}{\partial \lambda} \right) + \lambda \sum_{s=0}^{\infty} \frac{\partial^2 \log f_s}{\partial \lambda^2} + \lambda \left( \sum_{s=0}^{\infty} \frac{\partial \log f_s}{\partial \lambda} \right)^2 \right),$$

it suffices to show $(w + \sum_{s=0}^{\infty} \frac{\partial \log f_s}{\partial \lambda}) > 0)$. This is true because we have shown $\lim_{\lambda \to 0} (w + \sum_{s=0}^{\infty} \frac{\partial \log f_s}{\partial \lambda}) > 0) = 0$ and $\sum_{s=0}^{\infty} \frac{\partial^2 \log f_s}{\partial \lambda^2} > 0$.

This completes the proof that $\lambda^* < 0$ and hence we have completed the proof for this Lemma.